\documentclass{iau}

\usepackage{amsmath}
\usepackage{graphicx}
\usepackage{multirow}
\newcommand{\msun}{\mbox{$M_{\odot}$}}
\newcommand{\Msun}{\mbox{$M_{\odot}$}}

\begin{document}

\lefttitle{Cambridge Author}
\righttitle{Proceedings of the International Astronomical Union: \LaTeX\ Guidelines for~authors}

\jnlPage{1}{7}
\jnlDoiYr{2021}
\doival{10.1017/xxxxx}

\aopheadtitle{Proceedings IAU Symposium}
\editors{C. Sterken,  J. Hearnshaw \&  D. Valls-Gabaud, eds.}

\title{Very massive stars and Nitrogen-emitting galaxies}

\author{Jorick S. Vink}
\affiliation{Armagh Observatory and Planetarium}

\begin{abstract}
Recent studies of high-redshift galaxies using JWST, such as GN-z11 revealed highly elevated levels of nitrogen (N). This phenomenon extends to gravitationally-lensed galaxies like the Sunburst Arc at z = 2.37, as well as to globular clusters (GCs). We propose that this originates from the presence of very massive stars (VMSs) with masses ranging from 100 to 1000\,\Msun. The He {\sc ii} observed in the Sunburst Arc could also stem from the disproportionately large contribution of VMSs.
We build an entirely new Framework for massive star evolution which is no longer set by Dutch or other mass-loss "recipes" but which take the physics of $\Gamma$ or $L/M$-dependent winds into account. 
We discuss the mass-loss kink and the transition mass-loss rate between optically thin and thick winds, before
we study the evaporative mass-loss history of VMSs. Our novel evolution models exhibit vertical evolution in the HR-diagram from the zero-age main sequence due to a self-regulatory effect driven by their wind-dominated nature, and we discuss what wind physics sets the stellar upper-mass limit.  
Our estimate for the Sunburst Arc in \cite{V23} suggests that the significant amounts of N found in star-forming galaxies likely arise from VMSs.
We evaluate the strengths and weaknesses of previous hypotheses, including fast rotating massive stars and supermassive stars (SMSs), and we conclude that only our VMS model satisfies the relevant criteria.
Finally, we advocate for the inclusion of VMSs in population synthesis and chemical evolution models, emphasizing the need for a self-consistent wind approach, which currently does not exist. Even minor inaccuracies in mass-loss rates dramatically impact the stellar evolution of VMS, as well as their ionizing and chemical feedback.
\end{abstract}


\maketitle

\section{Introduction}

A central question in Astrophysics is whether there is an upper mass limit for stars and, if so, what physical mechanism determines this limit. The answer to this question could also indicate whether the upper mass limit depends on metallicity ($Z$). We proposed that mass loss due to radiation-driven winds influenced by line opacity is a key factor in determining the stellar upper mass limit \citep{V18}.

High-redshift galaxies are now commonly investigated using JWST. One surprising finding from galaxies like GN-z11 are the elevated nitrogen (N) abundances \citep{Cameron23,Sen23,Marques23,Pascale23}. A plausible explanation for this N enhancement is the production and subsequent stellar wind release of hydrogen (H) burning products through the CNO cycle.

Very massive stars (VMS) with masses in the range of 100-1000\,\Msun\ are formidable cosmic engines, as well as factories of chemical elements, such as N \citep{V23}, as well as sodium (Na), and aluminium (Al) relevant for globular clusters \citep{V18,H23}. 

\begin{figure}
    \includegraphics[scale=.4]{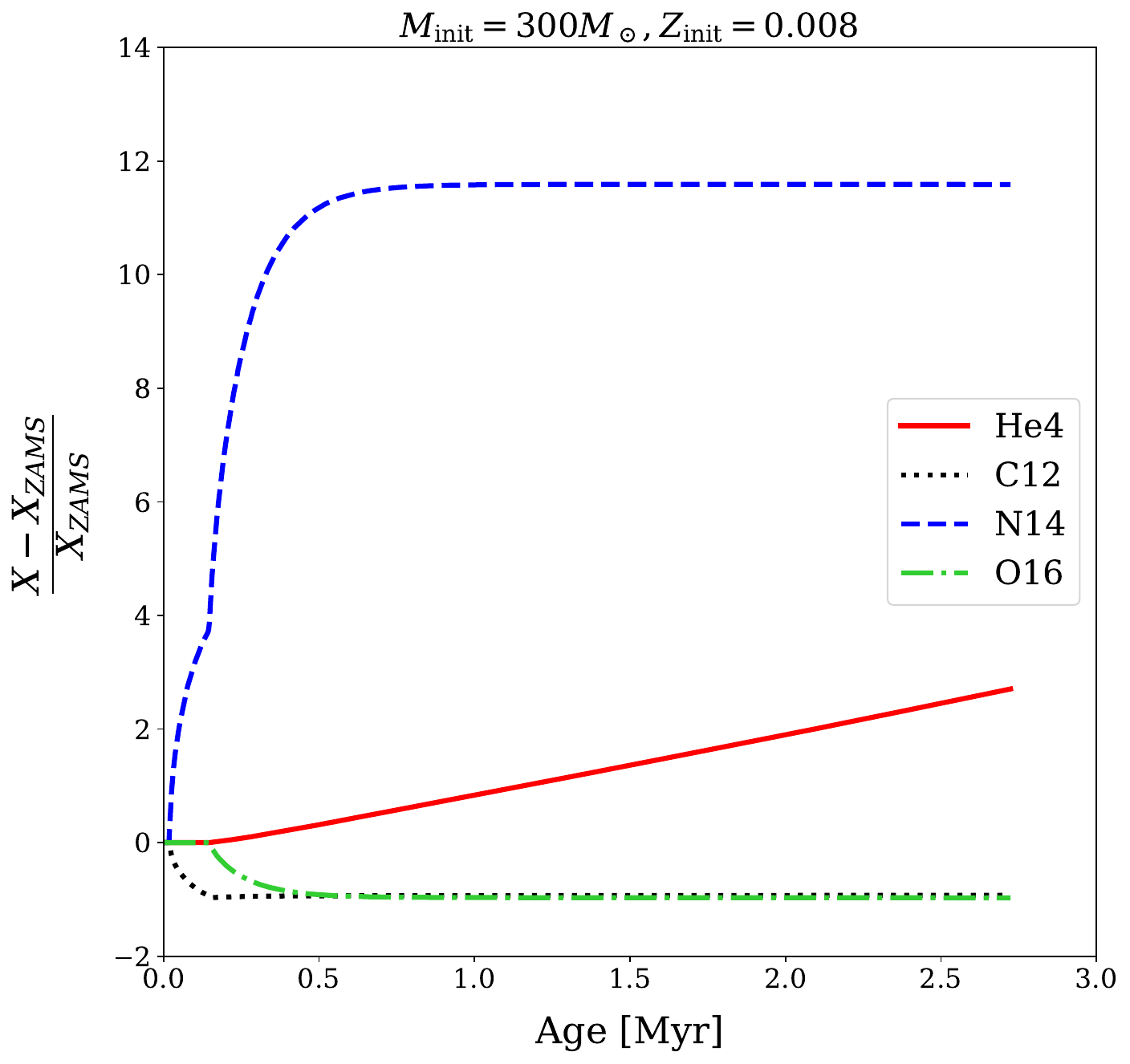}
    \caption{Relative surface enrichment for a 300\,$M_{\odot}$ VMS at sub-solar LMC-like metallicity. The VMS MESA stellar evolution models are from \cite{S22}.}
    \label{nitrogen}
\end{figure}

What makes VMS with masses in the 100 - 1000\msun\ range fundamentally different from canonical O stars in 20-100\,\msun\ range is their proximity to the Eddington $\Gamma$ limit. VMS have huge convective cores, and with enhanced mass-loss rates \citep{V11} they naturally undergo chemical homogeneous evolution (CHE) \citep{grafener11,Yusof13,S22} without the need to invoke CHE from fast stellar rotation \citep{VH17}.

\section{Alternative Polluters}

\subsection{Asymptotic Giant Branch stars}

While AGB stars are key producers of chemical elements in the Cosmos, it is unlikely that they are responsible for chemical enrichment of the Early Universe, as it takes a long evolutionary time for low and intermediate mass-stars to reach the AGB phase. In addition, the anti-correlations in GCs between Na and oxygen are not reproduced \citep{Bast18} For these reasons, massive stars appear to be the more likely culprits, but which massive stars?

\subsection{Fast rotating massive stars}

Another class of stellar objects that have been suggested to reproduce the anti-correlations in GCs are fast rotating massive stars \citep{Dec07}. 
However, this scenario would rely on a number of assumptions, such as high stellar rotation, end very efficient rotational mixing. Rotating stars leading to wind ejection in classical Wolf-Rayet (cWR) stars have also been invoked to explain the N in high-z galaxies \citep{Sen23}, but it should be realised that if cWR stars at lower $Z$ would indeed rotate more rapidly, the expectation would be that they should be more linearly polarised than their Galactic cWR
counterparts, which is not observed \citep{VH17}. It is thus questionable that fast rotating massive stars would be responsible for Na-O and Al-Mg anti-correlations in globular clusters (GCs), nor N-enhancement in high z star-forming galaxies.

\subsection{Supermassive stars}

Instead, supermassive stars (SMS, with $M \simeq 10^4 M_{\odot}$) \citep{Den14,Gieles18} have been proposed \citep{Char23} as potential contributors to the N enrichment in star-forming galaxies, as well as to the anti-correlations observed in globular clusters (GCs). 
One reason for this suggestion was because the winds from carbon-rich Wolf-Rayet stars (cWR) are too rapid to retain the enriched material within the gravitational influence of GCs.

A critical question that has not been addressed however is whether SMS exist in Nature.

\section{The role of VMS and SMS in the Universe}

\begin{figure}
    \includegraphics[scale=.5]{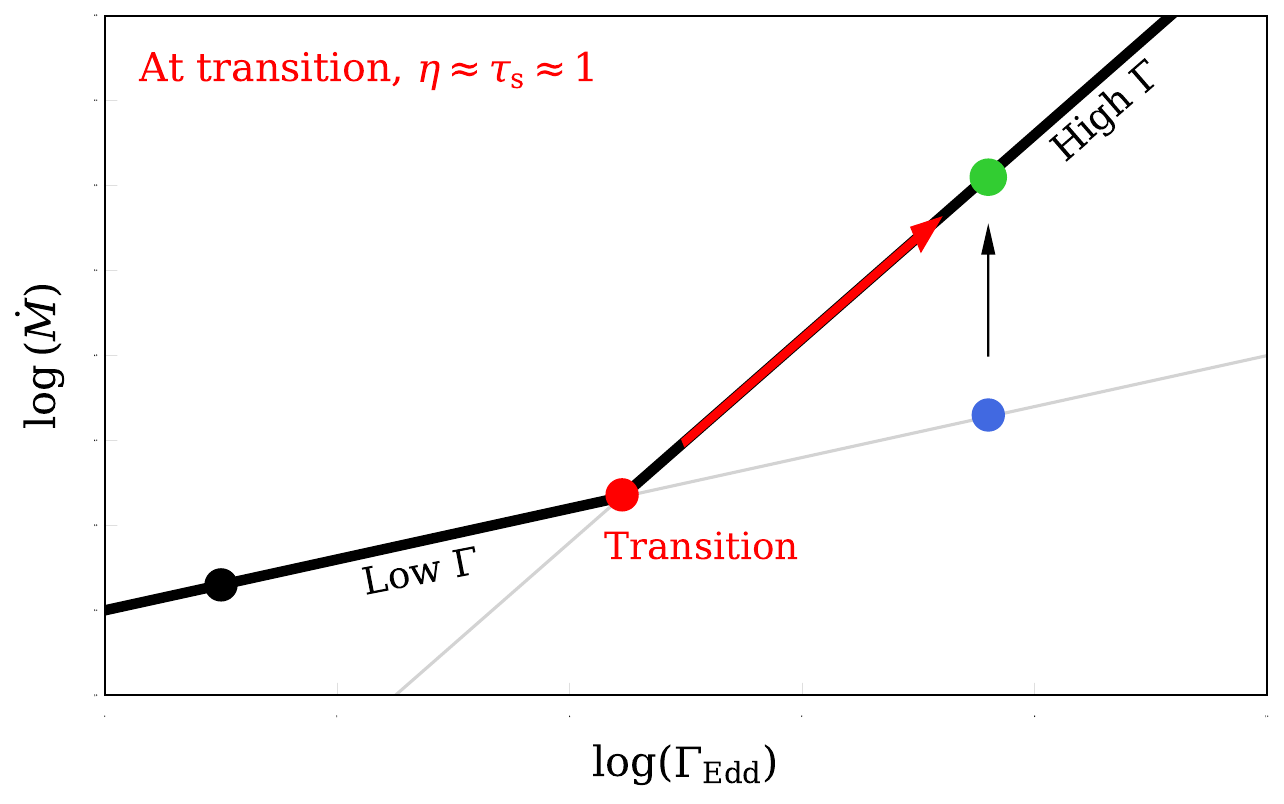}
\caption{A cartoon of the wind mass-loss rate versus the Eddington parameter $\Gamma$. The black dot represents stars in the traditional massive O-type regime, while the red dot provides the \cite{VG12} transition point. Above this red point, wind mass-loss rates need to be boosted from the blue to green dot, accounting for the much steeper \cite{V11} slope at higher $\Gamma$ (see \cite{S22}.}
\label{fig:kink}
\end{figure}

For many years, it had been believed that VMS with masses significantly exceeding 100\,\Msun\ may have been more common in the early Universe. This expectation stems from the idea that the first generations of stars, including pop {\sc iii} stars \citep{S02} were generally more massive due to reduced cooling during the formation of these metal-poor objects compared to today's metal-rich Universe \citep{Bromm99,Abel00}. 

But there is a second reason why VMS may be more common in the metal-poor Universe that that is due to the physics of radiation-driven winds that are thought to be weaker in the early Universe with its lower metal content, leading to a higher upper mass limit at lower $Z$ \citep{V18}. 
This could imply that the final masses of VMS would remain close to their initial masses, which could even lead to the creation of intermediate-mass black holes from VMS in the 100-1000\,\msun\ range, or even the direct creation of supermassive black holes around $10^5$ \msun\ found in galactic centers, if SMS exist.

\section{The very existence of VMS}

Prior to addressing the reality of SMS we should first recall the evidence for the existence of VMS.
\cite{Crow10} conducted a re-analysis of the most massive hydrogen- and nitrogen-rich Wolf-Rayet (WNh) stars located in the center of R136, the ionizing cluster of the Tarantula Nebula in the Large Magellanic Cloud (LMC). Their analysis revealed that stars previously thought to be below a "canonical" upper mass limit of 150\msun\ were actually more luminous, with initial masses reaching approximately 200-300\msun\ \citep{V15,MP22}. 

Some members of the astronomical community appear to remain skeptical about the existence of very high masses in R136, particularly in light of earlier extraordinary claims about SMS of thousands of solar masses in the Tarantula nebula in the 1980s, as  higher-resolution observations have revealed that R136 was not a single SMS but rather a young cluster containing several lower-mass objects, including the current record holder, R136a1.

For some it thus appears to remain questionable that the luminosities derived by Crowther et al. are overestimates, as the central WNh stars might actually involve multiple sources due to inadequate spatial resolution. 

However, while it would technically be possible that the most luminous stars in areas such as R136 could dissolve into smaller objects from a photometry point-of-view alone, the reality is that these WNh stars have very specific emission line spectra that can only be produced by VMS with extremely strong stellar winds \citep{V11}. By contrast, canonical O-type stars are predominately characterised by absorption lines.

As the VMS in R136 are in the emission-line dominated regime, breaking up one 300\msun\ star into say ten 30\Msun\ O-type stars cannot work, as that would lead to an integrated absorption line spectrum for R136, while He {\sc ii} is in fact observed in emission \citep{Crow16}. Let us examine the transition from optically thin ($\tau < 1$) O-type stars to optically thick ($\tau > 1$) VMS in the following.

\section{The extreme winds of VMS}

Canonical O-type stars stellar spectra that are characterised by absorption lines, but when stars become more luminous, and display increasingly larger mass-loss rates, He {\sc ii} absorption lines turn into emission \citep{V11,Best14}. This change-over happens when the wind optical depth $\tau$ and the wind efficiency number $\eta$ cross unity \citep{VG12}.

Figure \ref{fig:kink} shows a kink in the relationship between mass-loss rate and the Eddington $\Gamma$ parameter (which is proportional to the $L/M$ ratio).
The O-type stars on the lower slope of the relation have absorption lines, whilst those above the \cite{VG12} $\eta = \tau =1$ mass-loss transition "kink" have emission lines.

\begin{figure}
    \includegraphics[scale=.5]{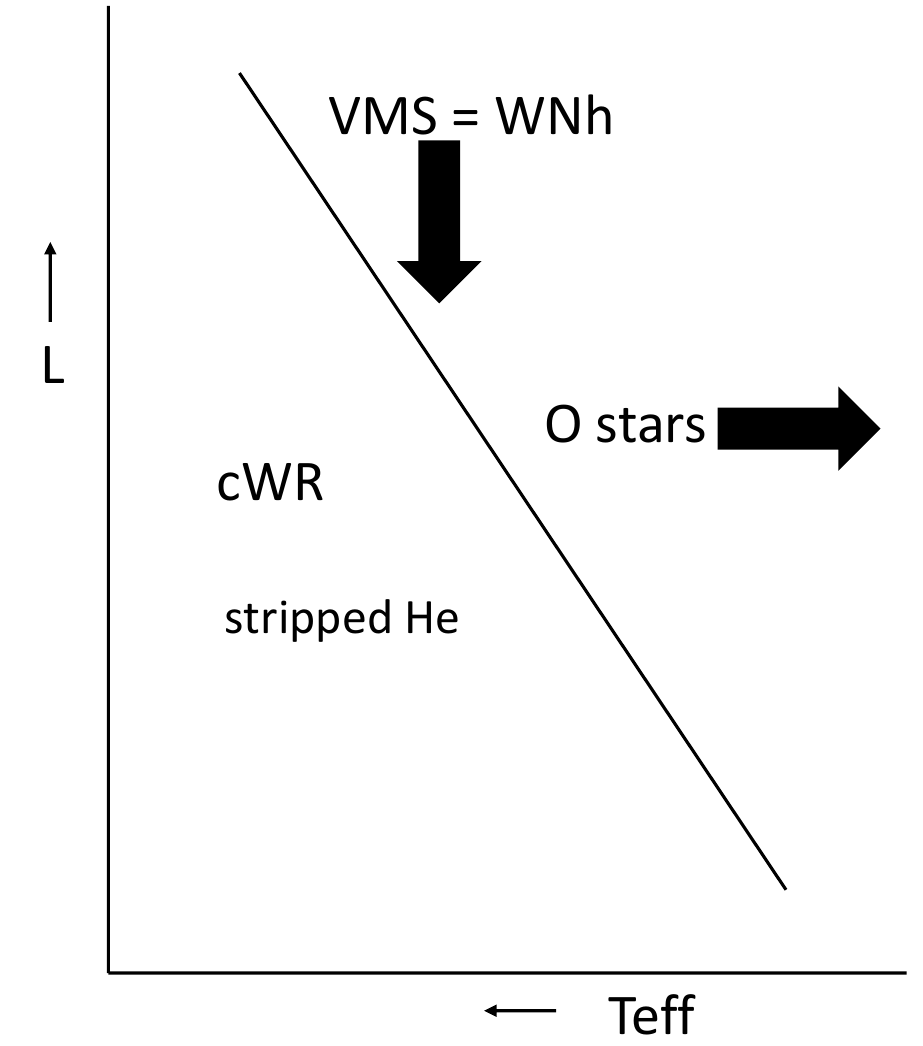}
    \caption{While canonical massive O-type stars are expected to evolve horizontally in the stellar HR diagram (HRD), very massive stars (VMS) are observed to evolve vertically downwards from the onset of hydrogen burning, as indicated by the downward thick arrow. These stars are likely to produce classical Wolf-Rayet stars (cWRs) located on the hot side of the slanted line representing the zero-age main sequence (ZAMS). At luminosities below a certain threshold \citep{SV20}, these stars may appear as optically thin stripped helium stars.}
    \label{HRD}
  \end{figure}

In addition to affecting the integrated spectra of star-forming galaxies \citep{SB99,Mestric23}, the stronger mass-loss slope of VMS also has a decisive effect on the stellar evolution, the final black hole (BH) mass, and the chemical yields from strong stellar winds.

For this reason, in Armagh we have recently implemented the $\eta = \tau =1$ mass-loss transition strategy into the MESA \citep{Pax13} stellar evolution code \citep{S22,S23}, and we 
encounter vertical evolution in the HR diagram, as opposed to traditional horizontal evolution (see Figure.\,\ref{HRD}). 
The enhanced wind mass loss also leads to a mass-evaporation effects, with consequences for the upper mass limit to stars \citep{V18}.

\section{Mass Evaporation and the Hydrogen Clock}

\begin{figure}
    \includegraphics[scale=.45]{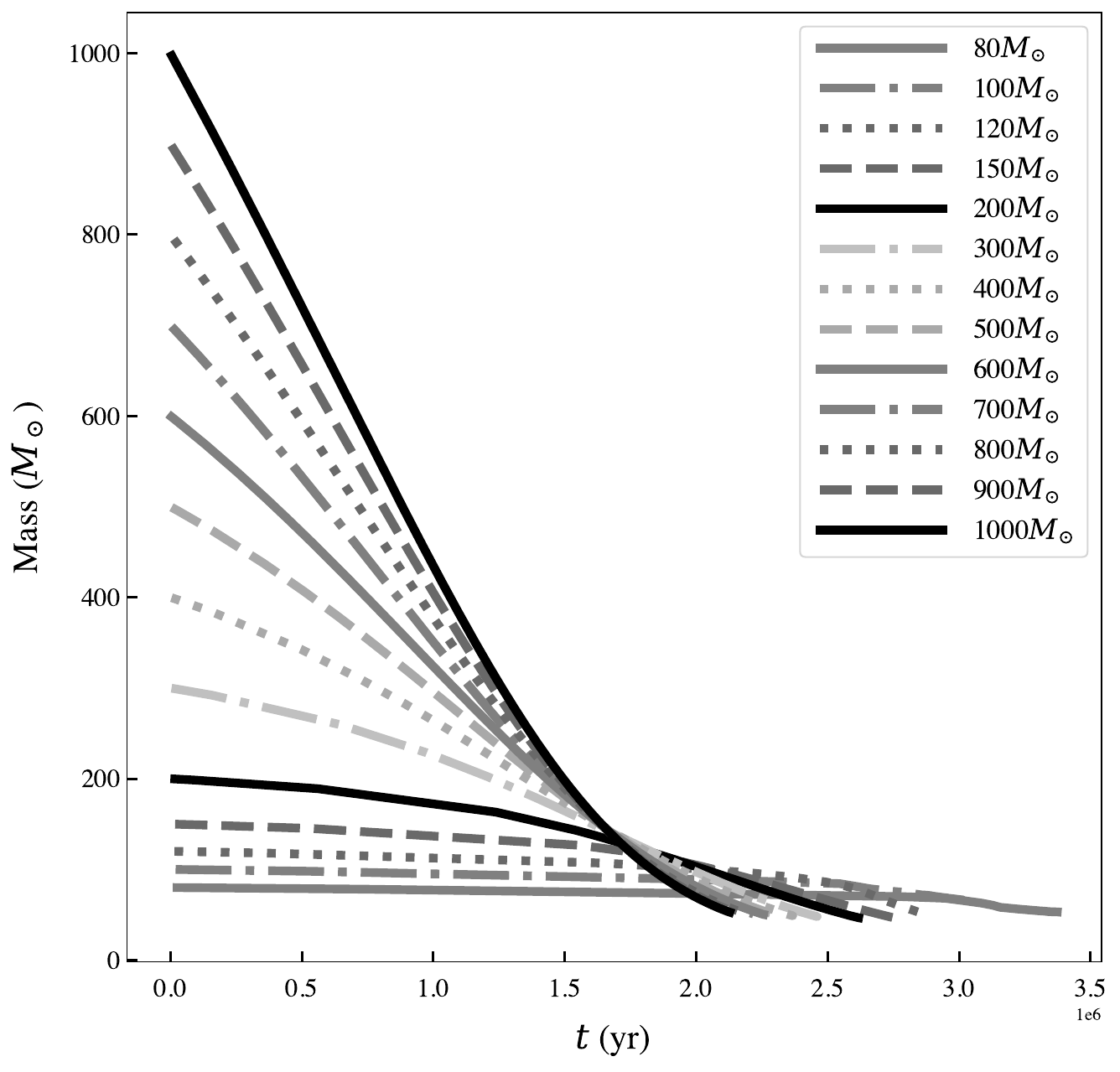}
    \caption{The "effective upper mass limit" and stellar mass evolution of VMS up to 1000\,\msun, as they progress from Zero Age Main Sequence (ZAMS) to Terminal Age Main Sequence (TAMS). The alternating gray lines indicate the different masses within the grid. 
    The figure emphasizes that it is not feasible to determine the initial upper mass limit of stars based on present-day observations, unless one knows the precise mass-loss history \citep{V18,S22,H22}.}
    \label{clock}
  \end{figure}
  
When we include the enhanced winds into stellar evolution models, we obtain a strong drop of the stellar mass with time, as shown in Figure.\,\ref{clock}. In particular, a convergence to a similar stellar mass is observed around 1.5 million years -- regardless of wether the initial mass was say 300\,\msun\ or 1000\,\msun.

Furthermore, the surface hydrogen abundance $X_{\rm S}$ can serve as the key for determining initial mass, as VMS evolve chemically homogeneous above a certain mass threshold. 
In \cite{H22} we utilized this tool to resolve the ambiguity in the initial masses of both components in a detached binary system as well as amongst a larger sample of WNh stars in the Tarantula Nebula. For some objects we found that the initial stellar mass is completely unrestricted, with options easily extending to 1000\Msun. 

In short, the mass turnover point at 1.5 million years in combination with $X_{\rm S}$ can function as a 'clock' for estimating the stellar upper mass limit.

\section{Summary}

\begin{itemize}
    
    \item
VMS in the 100 - 1000\,\msun\ range are {\it fundamentally} different from canonical O stars in the 20-100\msun\ range. For this reason, simply discussing a potentially "top-heavy" nature of the IMF of high-redshift star forming galaxies has very limited value. 

\item 
VMS produce the He {\sc ii} emission from young stellar clusters in the Local Universe.

\item
VMS may produce excessive amounts of N, and could be responsible for the exciting JWST observations of galaxies such as GN z-11. Additional "by-products" from H-burning via the CNO cycle, such as Na and Al, could be relevant for the self-enrichment of globular clusters.

\end{itemize}

\end{document}